\documentclass{article}

\usepackage{arxiv}

\usepackage[utf8]{inputenc} 
\usepackage[T1]{fontenc}    
\usepackage{hyperref}       
\usepackage{url}            
\usepackage{booktabs}       
\usepackage{amsfonts}       
\usepackage{nicefrac}       
\usepackage{microtype}      
\usepackage{lipsum}
\usepackage{graphicx}
\usepackage{amsmath,amssymb,amsfonts}%

\usepackage[math]{cellspace}
\title{Hierarchical Single-Linkage Clustering for Community Detection with Overlaps and Outliers}
\author{
 Ryan DeWolfe \\
  Department of Mathematics\\
  Toronto Metropolitan University\\
  Toronto, Canada \\
  \texttt{ryan.dewolfe@torontomu.ca} \\
}

\begin{document}
\maketitle
\begin{abstract}
    Most community detection approaches make very strong assumptions about communities in the data, such as every vertex must belong to exactly one community (the communities form a partition).
    For vector data, \textit{Hierarchical Density Based Spatial Clustering for Applications with Noise} (HDBSCAN) has emerged as a leading clustering algorithm that allows for outlier points that do not belong to any cluster.
    The first step in HDBSCAN is to redefine the distance between vectors in such a way that single-linkage clustering is effective and robust to noise.
    Many community detection algorithms start with a similar step that attempts to increase the weight of edges between similar nodes and decrease weights of noisy edges.
    In this paper, we apply the hierarchical single-linkage clustering  algorithm from HDBSCAN to a variety of node/edge similarity scores to see if there is an algorithm that can effectively detect clusters while allowing for outliers.
    In experiments on synthetic and real world data sets, we find that no single method is optimal for every type of graph, but the admirable performance indicates that hierarchical single-linkage clustering is a viable paradigm for graph clustering.
\end{abstract}

\section{Introduction}\label{sec1}
Finding groups of similar data in an unsupervised method, called clustering, is a fundamental problem in data science \cite{clusteringhandbook}.
When working with data in the form of a graph, we often consider an edge as an indicator of similarity between two nodes, and clustering (or community detection) involves finding sets of nodes that have many edges between them.
Despite the incomplete definition, clustering graphs is an important process with many applications, including link prediction, event detection, and biology \cite{fortunato_review}.

Many graph clustering algorithms have been proposed \cite{fortunato_review}, but unfortunately most of them can only return a partition of the data.
Some papers \cite{linkcoms,egosplit,linegraphclustering} have argued for overlapping partitions, but very few algorithms allow for clusterings that do not necessarily cover all the nodes.
In fact, the terms partition and clustering are often used interchangeably; we make the distinction that a clustering does not necessarily cover all the nodes, so that a partition is a clustering but a clustering may not be a partition.

In this paper, we apply the \textit{Hierarchical Single-Linkage Clustering} (HSLC) algorithm from HDBSCAN \cite{hdbscan} to the undirected weighted graphs produced by several previously proposed node or edge similarity scores \cite{linkcoms,berry_2011,unweighted_linegraphclustering,node2vec,simrank,rw_weighting,nbrw2021,ecg2} in order to create a variety of unsupervised graph clustering algorithms.
HSLC is particularly attractive since it has only a single parameter (the minimum cluster size), and it makes few assumptions about the clusters in the data \cite{hdbscanalg}.
When we apply HSLC to edge similarities, the edge clusters can be projected to the nodes, and the result is a clustering that can have both outliers and overlapping clusters.

\section{Node and Edge Similarity Measures}
The first step in the HDBSCAN \cite{hdbscan} algorithm constructs a new notion of dissimilarity that is robust to noise in the data.
Many graph clustering/partitioning algorithms start with a similar step that attempts to down-weight edges that are believed to be inter-cluster and up-weight edges that are believed to be intra-cluster.
This is usually followed by running a standard partitioning algorithm on the weighted graph, since a good weighting of the edges has been shown to help lessen the resolution limit issue of modularity \cite{berry_2011}.
In this section, we review several proposed methods for evaluating the similarity between nodes/edges, represented by a function $s$, where a higher edge weight means more similarity.
Some methods define similarity between any pair of nodes/edges, while others only provide weights for adjacent nodes/edges, so in our experiments we restrict ourselves to using the similarity measures of adjacent nodes/edges (and in this sense we are applying a weighting to the graph/line graph).

\subsection{Node Similarity Measures}

\subsubsection{Short Cycles (SC) \cite{berry_2011}}
The intuition for this method is that short cycles are more prevalent within clusters than between clusters.
They propose weighting the edges of the graph based on the number of triangles and rectangles the edge is in, normalized by the maximum possible number of triangles and rectangles given the degrees of the ends.
Furthermore, they show that iterating the weighing process improves the results (we use $3$ iterations).

\subsubsection{Random Walk Weighting (RWW) \cite{rw_weighting}}
This method uses short random walks to quantify the similarity between nodes, following the intuition that a random walk is more likely to stay within a community \cite{rw_weighting,walktrap}.
This method considers random walks up to length $\ell$ (with experiments for $\ell=3,4$ in the paper) starting at a source node $n$, and creates a vector corresponding to the probability of ending a random walk on each vertex.
This is computed by creating a transition matrix $T$, equivalent to a row-normalized adjacency matrix, and then computing
$$ P = \sum_{x=1}^\ell T^x $$
The weight of an edge $s(i,j)$ is defined as the cosine similarity between rows $i$ and $j$ of $P$.
The weighting process is run iteratively to get a final weighting of the network (we use $3$ iterations and $\ell = 3$).

\subsubsection{Node2vec (N2V) \cite{node2vec}}
The node2vec algorithm uses many samples of short random walks starting at each node to embed the nodes of a graph into a vector space $\mathbb{R}^d$, and these embeddings have been shown to work well for graph partitioning \cite{n2v_comdetection}.
For the weight of an edge $e_{ij}$,  we use $s(i,j) = \frac{1}{1+||v(i) - v(j)||_2},$ where $v(i)$ and $v(j)$ are the node2vec embeddings of nodes $i$ and $j$ respectively.
We the use pecanpy implementation \cite{pecanpy} with parameters: $p=q=1$, $40$ walks per node, $80$ steps per walks, and $d=16$.

\subsubsection{Renewal Non-backtracking Random Walks (RNBRW) \cite{nbrw2021}}
This paper combines the intuitions of cycles and random walks.
A non-backtracking random walker starts at a random vertex and walks until it creates a cycle, which is equivalent to revisiting a vertex since the walk in non-backtracking (walks that get stuck are discarded).
Each edge is weighted with the probability that it is the last edge traversed in this random walk process, and ``edges with larger weights may be thought of as more important to the formation of cycles'' \cite{nbrw2021}.
In practice, the edge weights are computed by sampling a large number of random walks, which is set to $m$ (the number of edges in the graph) following the default implementation.

\subsubsection{SimRank \cite{simrank_weighting}}
Zhang et al.\ propose a combination of the common neighbor index and Simrank \cite{simrank} that allows similarity to propagate beyond the immediate neighborhood of each node.
Let $s_t(i,j)$ represent the similarity between nodes $i$ and $j$ in iteration $t$.
Initialize $s_0(i,j)$ with the indicator function $\chi\{i = j\}$, and define the similarity in round $t$ as
$$ s_t(i,j) = \frac{1}{|N(i)| \times |N(j)|} \sum_{x \in N(i)} \sum_{y \in N(j)} s_{t-1}(x, y).$$
In experiments on LFR and real world data, the authors found optimal $t$ values between $1$ and $5$, so we take $t=3$ to match the other iterative methods.

\subsubsection{Ensemble Clustering for Graphs (ECG) \cite{ecg2}}
ECG was developed to address stability concerns with the Louvain partitioning algorithm.
Many independent runs of the first stage of Louvain are run, and the edges of the graph are weighting according to the fraction of times the endpoints end up in the same part.
Due to the randomized implementation of Louvain, this method is not deterministic, although the original paper found stable results for an ensemble size of $16$ (which we also use here).

\subsection{Edge Similarity Measures}

\subsubsection{Link Communities (LC) \cite{linkcoms}}
This methods evaluates the similarity of two adjacent edges based on the neighborhood overlap of the nodes at either end of the two-path.
Let $N[i]$ be the closed neighborhood of node $i$ (all of $i$'s neighbors and $i$ itself).
Then, for edges $e_{ij}$ and $e_{jk}$, the similarity is defined as
$$s(e_{ij}, e_{jk}) = \frac{\left|N[i] \cap N[k]\right|}{\left|N[i] \cup N[j]\right|}.$$

\subsubsection{Line Graph Transition Probabilities (LGTP) \cite{unweighted_linegraphclustering}}
Evans and Lambiotte propose weighting the line graph according to the transition probabilities of an edge-based random walk.
A random walker starts on an edge, and moves to an adjacent edge by first selecting one of the endpoints with equal probability and then selecting a random edge incident to that endpoint with probabilities proportional to the weight of each edge.
Thus, for any adjacent edges $e_{ij}, e_{jk}$, $s(e_{ij}, e_{jk}) = \frac{1}{deg(j)}$.

\subsubsection{Node Similarity Measures on the Line Graph (LG-\textit{method})}
These methods have not been explicitly proposed previously, but inspired by \cite{unweighted_linegraphclustering}, any of the node similarity scores can be run on the line graph to obtain an edge similarity score.
We use the line graph weighted by transition probabilities (see the previous method) to down-weight the contribution of large degree nodes, since they produce a large cliques in the line graph.
Furthermore, the line graph is often much larger than the original graph so there is concern about the scalability of these methods.
We found RWW and Simrank are too slow, and reduced the number of walks per node to $10$ and the steps per walk to $20$ for node2vec. 

\subsubsection{Edge ECG (EECG)}
Finally, we propose same idea as ECG \cite{ecg2}, except instead of weighting adjacent nodes, adjacent edges are weighted as the proportion that all three nodes are in the same part after the first stage of Louvain.

\section{Hierarchical Single-Linkage Clustering}\label{sec:hdbscan_review}
In this section, we describe the HSLC clustering algorithm as it is applied to the similarity graphs described in the previous section.
HDBSCAN \cite{hdbscan} has strong theoretical foundations for vector data, but unfortunately they do not transfer to similarity graphs so we refer the interested reader to \cite{hdbscanalg} for the motivation behind the algorithm.
We note that the algorithm was originally written for dissimilarity scores (distances) in \cite{hdbscan,hdbscanalg}, so we invert many of the definitions to continue with a similarity perspective.
Finally, since we have defined similarity between both nodes and edges, we define the clustering method on a generic similarity graph $S$, with vertices $K$ and weighted edges $L$.

First, a single-linkage dendrogram is built from the similarity graph $S$.
The clusters at level $\lambda$ are the connected components of the subgraph with all vertices and edges $\{k_1k_2 \in L : s(k_1, k_2) \geq \lambda\}$.

However, this dendrogram is generally too complex to visualize (consider that when $\lambda \to \infty$ every object has its own cluster).
To condense the dendrogram so that it only tracks significant clusters, a parameter $m_s$ is introduced to control the minimum cluster size.
This does impose a type of \textit{resolution limit} \cite{berry_2011} to the algorithm, but the effect of this parameter is intuitive.
When traversing the dendrogram top-down, if a cluster is split, then one of the following three cases will occur:
\begin{enumerate}
    \item The cluster splits into several clusters, each with less than $m_s$ vertices. We say the cluster has disappeared.
    \item The cluster splits into several clusters, at least two of which have at least $m_s$ vertices. We say all sub-clusters with at least $m_s$ vertices are significant and different from the original.
    \item The cluster splits into one cluster with at least than $m_s$ vertices and one or more clusters with less than $m_s$ vertices. We say the original cluster shrinks, and the largest sub-cluster retains the name of the original cluster. The other clusters become noise.
\end{enumerate}
The condensed dendrogram will have far fewer clusters, and naturally defines a level at which every cluster will disappear (for $m_s > 1$).

Finally, even though the condensed dendrogram is practical for investigation, many applications still require a single set of clusters.
HDBSCAN \cite{hdbscan} defines a persistence score for each cluster, and then provides an algorithm to find the set of non-overlapping clusters to maximize total persistence.
For a cluster $C_i \subseteq K$, define the death of the cluster as $\lambda_{min}(C_i) = \min\{\lambda: C_i \text{ exists} \}$.
Also, define the contribution of each object $k_j \in C_i$ as $\lambda_{max}(k_j, C_i) = \max\{\lambda : k_j \in C_i\}$.
Then, the persistence of cluster $C_i$ is given by the equation
$$ \sigma(C_i) = \sum_{k_j \in C_i} \left(\lambda_{max}(k_j, C_i) - \lambda_{min}(C_i)\right).$$
The optimal flat clustering is described as the set of clusters $\mathcal{C}$ that maximizes $\sum_{C_i \in \mathcal{C}} \sigma(C_i)$ subject to $C_i \cap C_j = \emptyset \;\forall\; C_i, C_j \in \mathcal{C}$.

If $S$ is a similarity graph of the nodes, HSLC returns a set of non-overlapping clusters while allowing for outliers.
However, if $S$ is a similarity graph of the edges, we project the non-overlapping edge clusters found by HSLC to the nodes by including a node in a cluster if it has at least on edge in the cluster.
The clustering of nodes induced by the clustering of edges can have both outliers and overlapping clusters.

\section{Results}
In the most general setting, both the prediction and the labels can have overlap and outliers.
We follow \cite{egosplit} and use precision, recall, and F1 score, although we report weighted averages since the distribution of community sizes is often far from uniform.
For a single cluster $C \subseteq N$ and a single label $L \subseteq N$, define the precision as $p(C, L) = |C \cap L| \slash |C|$, the recall as $r(C, L) = |C \cap L| \slash |L|$, and the F1 score as $\mathrm{F1}(C, L) = 2 p(C, L) \times r(C, L) \slash (p(C, L) + r(C, L))$.
However, there is not a matching of predicted clusters to labels, so the predicted cluster is compared to each label and the best is chosen.
Finally, a weighted average is used to combine the scores of each predicted cluster, with each cluster contributing proportional to its size.
Let $\mathcal{C}$ be a set of predicted clusters, and $\mathcal{L}$ a set of labels.
The weighted average scores are defined as
\begin{align*}
    \bar{p}(\mathcal{C}, \mathcal{L}) = \frac{1}{\sum_{C \in \mathcal{C}} |C|} \sum_{C \in \mathcal{C}} \left(|C| \times \max_{L \in \mathcal{L}}\{p(C, L)\} \right),\\
    \bar{r}(\mathcal{C}, \mathcal{L}) = \frac{1}{\sum_{C \in \mathcal{C}} |C|} \sum_{C \in \mathcal{C}} \left(|C| \times \max_{L \in \mathcal{L}}\{r(C, L)\} \right),\\
    \bar{\mathrm{F1}}(\mathcal{C}, \mathcal{L}) = \frac{1}{\sum_{C \in \mathcal{C}} |C|} \sum_{C \in \mathcal{C}} \left(|C| \times \max_{L \in \mathcal{L}}\{\mathrm{F1}(C, L)\} \right).
\end{align*}
The precision is the most important measure if our goal is a conservative algorithm that makes few mistakes.
We also report the coverage of a clusters (the percentage of objects with at least one cluster) since it is an intuitive value that is not obvious from the three measures above. 

For testing the clustering algorithms, we use synthetic graphs and real world data with known ground truth communities.
The synthetic graphs are generated from extensions of the \textit{Artificial Benchmark for Community Detection} (ABCD) \cite{abcd}, which is similar to the very popular LFR model, although the noise parameter $\xi \in [0,1]$ allows for a smooth transition from disjoint communities when $\xi=0$ to no community structure when $\xi=1$.
For non-overlapping node clustering, we use an extension that includes outlier nodes, ABCD+o \cite{abcdo}, with various proportions of outlier nodes.
For overlapping clusters, we use ABCD+o$^2$ \cite{abcdoo} graphs that have both outliers and overlapping ground truth clusters.

Finally, we consider four real data sets from various domains with known ground truth communities.
First, we use the Football graph \cite{football}, which has non-overlapping node clusters and known anomalous teams that we label as outliers \cite{football_coms}.
Next, we use a union K-nearest-neighbors graph from the MNIST digits dataset \cite{mnist}, with $10$ ground truth labels corresponding to each digit.
We set $K=15$, and there is an edge $e_{ij}$ if either $i$ in $j$'s 15 nearest neighbors or $j$ is in $i$'s 15 nearest neighbors.
K-nearest-neighbors graphs are commonly used for non-linear dimension reduction techniques, such as UMAP \cite{umap}, which are able to separate the clusters in the low dimensional output.
Finally, we use the DBLP (academic collaboration) and Amazon (co-purchasing) networks from the SNAP repository \cite{SNAP}, which have both outliers and overlapping cluster labels.

\subsection{Node Clustering}
\begin{figure}[t]
    \centering
    \large{10\% Outliers}\\
    \includegraphics[width=0.495\linewidth]{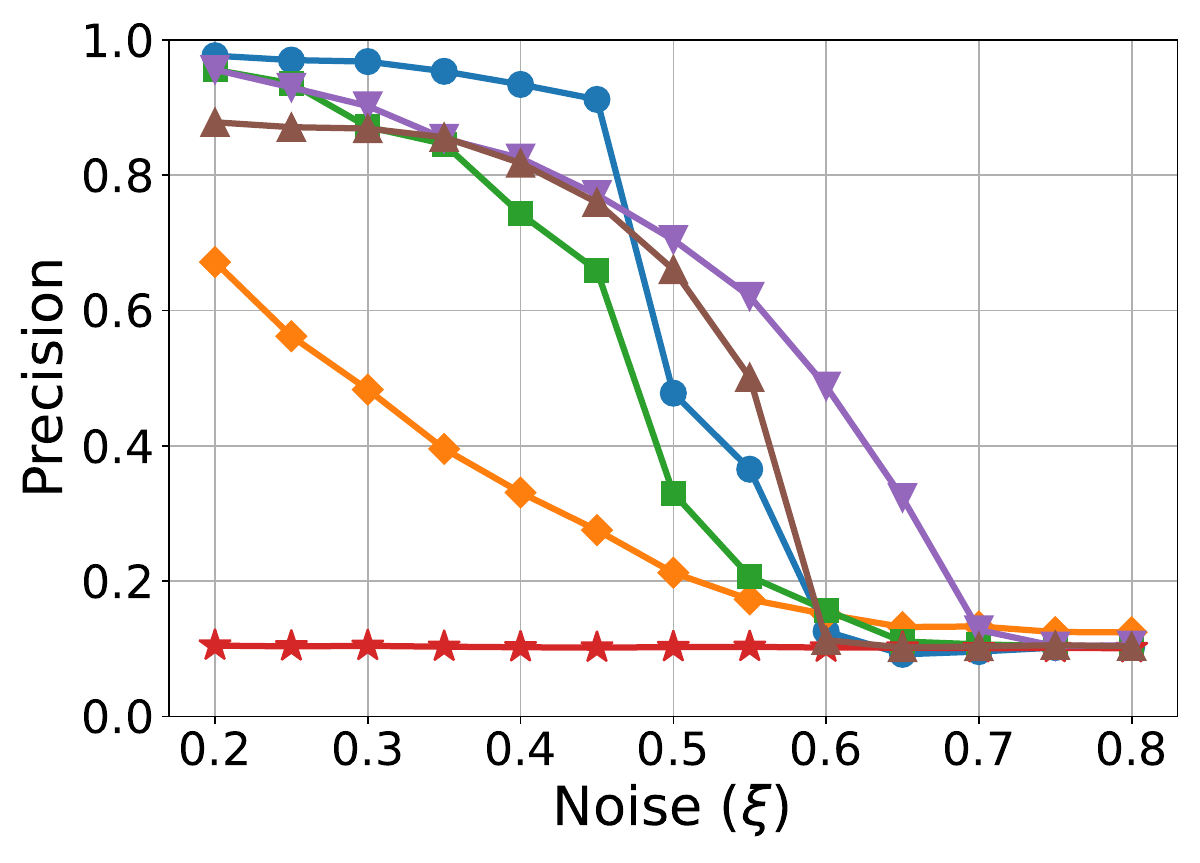}
    \includegraphics[width=0.495\linewidth]{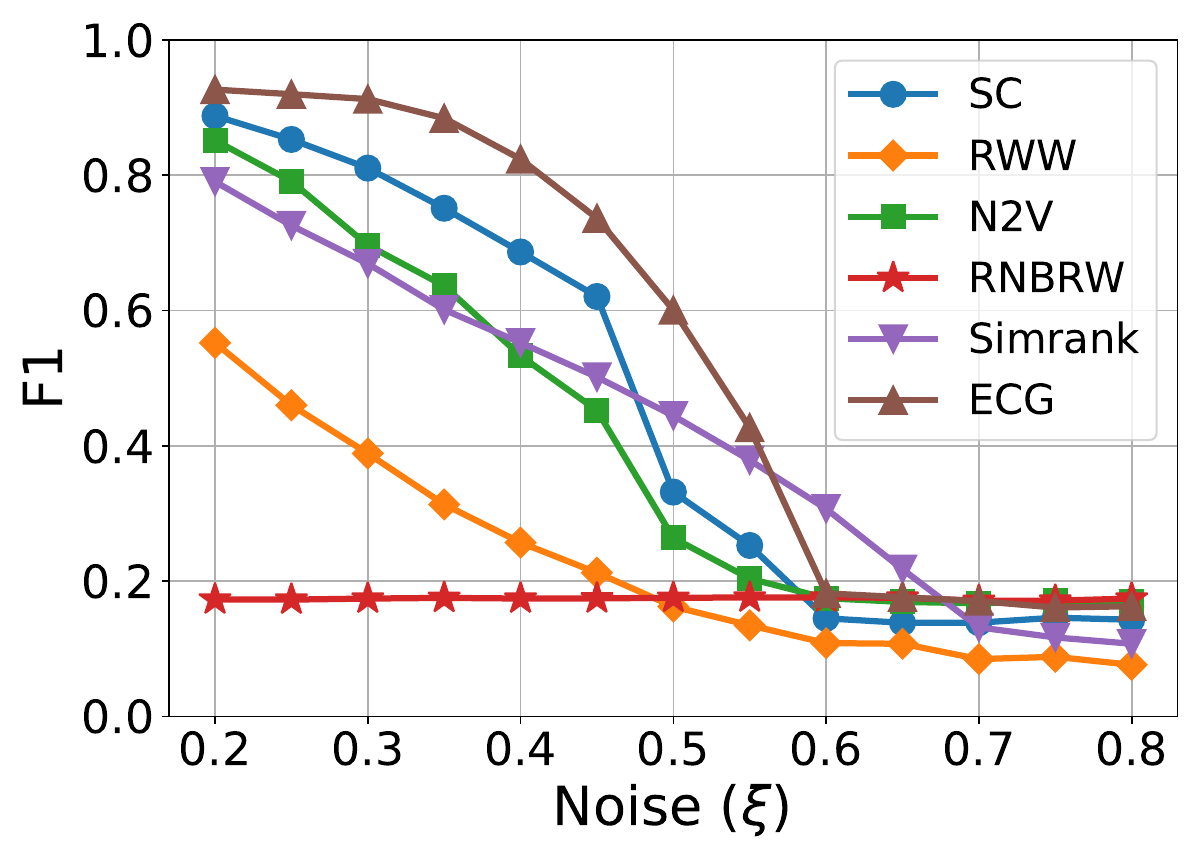}\\
    \large{50\% Outliers}\\
    \includegraphics[width=0.495\linewidth]{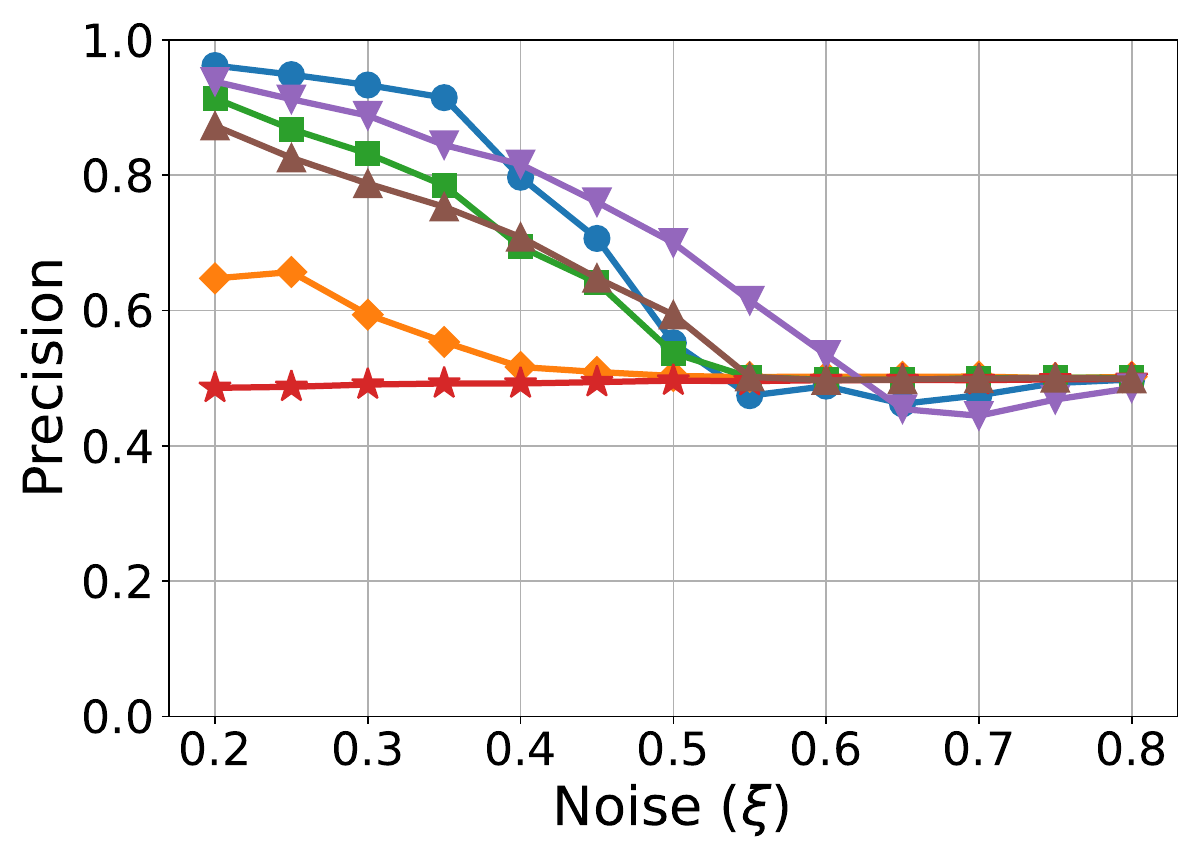}
    \includegraphics[width=0.495\linewidth]{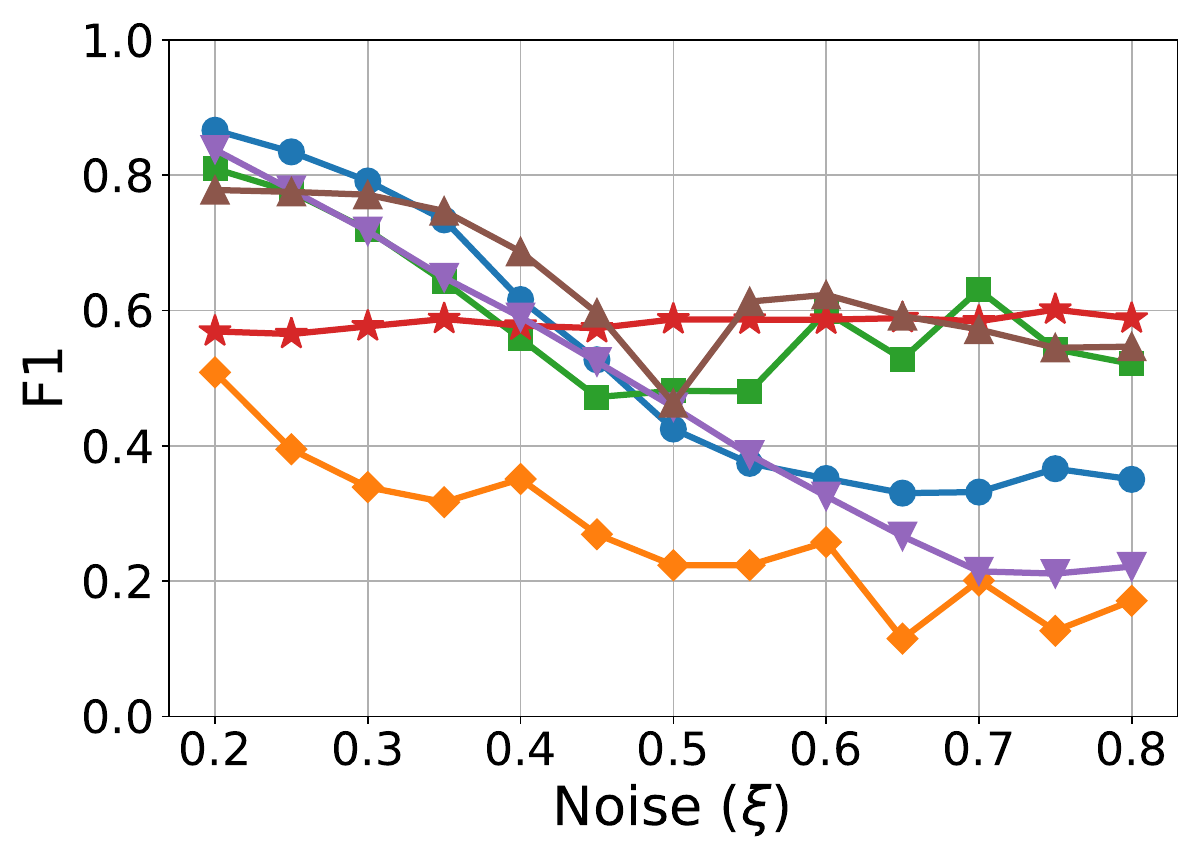}\\
    \caption{
        Results of the node clustering benchmark on ABCD+o graphs with $m_s = 15$.
        We report the precision (left) and F1 score (right) on ABCD+o graphs with $10,000$ nodes with 10\% outliers (top) or 50\% outliers (bottom).
        In each figure, we vary the noise parameter $\xi$ from $0.2$ to $0.8$, and report the average metric across $25$ random graphs.
    }
    \label{fig:abcdo_benchmark}
\end{figure}

First, we run an experiment on synthetic ABCD+o \cite{abcdo} graphs with $10,000$ nodes, one of 10\% or 50\% outliers, and a varying level of noise.
In Figure~\ref{fig:abcdo_benchmark}, we report the $\bar{p}$ and $\bar{\mathrm{F1}}$ versus $\xi$ curves of each node similarity method combined with HSLC.
No single weighting method performs best across metrics, proportion of outliers, or noise level.
ECG consistently achieves the best or competitive $\bar{\mathrm{F1}}$ score up to $\xi \approx 0.6$, at which point the graph is very noisy and no method is performing well.
With low noise values ($\xi \leq 0.4$), the short cycle method (SC) achieves the best precision, although ECG and Simrank are competitive, and perform better in the medium noise regime $\xi \in (0.4, 0.6]$.

Next, we apply each of the node clustering methods to the real world data-sets, setting $m_s=5$ for Football, $m_s = 500$ for MNIST, and $m_s = 10$ for DBLP and Amazon.
Results are shown in Table~\ref{tab:node_clustering_results}.
Similar to the experiment on synthetic graphs, no method clearly outperforms the others.
Node2vec (N2V), Simrank, Random walk weighting (RWW), and ECG perform fairly well for each graph.

\begin{table}[ph!]
    \centering
    \caption{
        Results from HDBSCAN cluster selection run on the node similarity graphs.
        We set $m_s$ to $5$, $500$, $10$, $10$ for the Football, MNIST, DBLP and Amazon graphs respectively.
        The largest precision and F1 score for each graph has been bolded.
        }
    \vspace{1em}
    \begin{tabular}{|Sc|Sc|Sc|Sc|Sc|Sc|} \hline
        Method & \# Clusters & Max Cluster Size & Coverage & Precision & F1 \\
        \hline
        \hline
        \multicolumn{6}{|Sc|}{Football} \\
        \hline
        SC & $3$ & $93$ & $0.98$ & $0.29$ & $0.38$ \\
        N2V & $11$ & $15$ & $0.97$ & $\mathbf{0.91}$ & $\mathbf{0.92}$ \\
        ECG & $10$ & $16$ & $1.00$ & $0.87$ & $0.91$ \\
        RNBRW & $8$ & $25$ & $0.78$ & $0.59$ & $0.61$ \\
        SIMRANK & $11$ & $16$ & $1.00$ & $0.88$ & $0.90$ \\
        RWW & $11$ & $19$ & $1.00$ & $0.88$ & $0.90$ \\
        \hline
        \hline
        \multicolumn{6}{|Sc|}{MNIST} \\
        \hline
        SC & $11$ & $7482$ & $0.37$ & $0.86$ & $0.59$ \\
        N2V & $9$ & $13865$ & $0.67$ & $0.80$ & $0.70$ \\
        ECG & $10$ & $7963$ & $0.99$ & $\mathbf{0.97}$ & $\mathbf{0.97}$ \\
        RNBRW & $2$ & $48498$ & $0.71$ & $0.13$ & $0.21$ \\
        SIMRANK & $3$ & $8171$ & $0.14$ & $0.77$ & $0.69$ \\
        RWW & $15$ & $7509$ & $0.44$ & $0.83$ & $0.51$ \\
        \hline
        \hline
        \multicolumn{6}{|Sc|}{DBLP} \\
        \hline
        SC & $5540$ & $1469$ & $0.37$ & $\mathbf{0.53}$ & $\mathbf{0.32}$ \\
        N2V & $9698$ & $323$ & $0.58$ & $0.49$ & $0.28$ \\
        ECG & $4894$ & $16142$ & $0.93$ & $0.28$ & $0.22$ \\
        RNBRW & $225$ & $300580$ & $0.96$ & $0.03$ & $0.05$ \\
        SIMRANK & $10385$ & $115$ & $0.62$ & $0.46$ & $0.30$ \\
        RWW & $7501$ & $144$ & $0.44$ & $0.51$ & $0.31$ \\
        \hline
        \hline
        \multicolumn{6}{|Sc|}{Amazon} \\
        \hline
        SC & $8296$ & $556$ & $0.63$ & $\mathbf{0.87}$ & $0.52$ \\
        N2V & $9971$ & $390$ & $0.65$ & $\mathbf{0.87}$ & $0.51$ \\
        ECG & $2452$ & $9471$ & $0.96$ & $0.76$ & $0.50$ \\
        RNBRW & $597$ & $307076$ & $0.95$ & $0.18$ & $0.28$ \\
        SIMRANK & $10507$ & $147$ & $0.61$ & $0.85$ & $\mathbf{0.53}$ \\
        RWW & $8678$ & $379$ & $0.51$ & $\mathbf{0.87}$ & $0.49$ \\
        \hline
    \end{tabular}
    \label{tab:node_clustering_results}
\end{table}

\subsection{Edge Clustering / Overlapping Clustering}
\begin{figure}[t]
    \centering
    \includegraphics[width=0.495\linewidth]{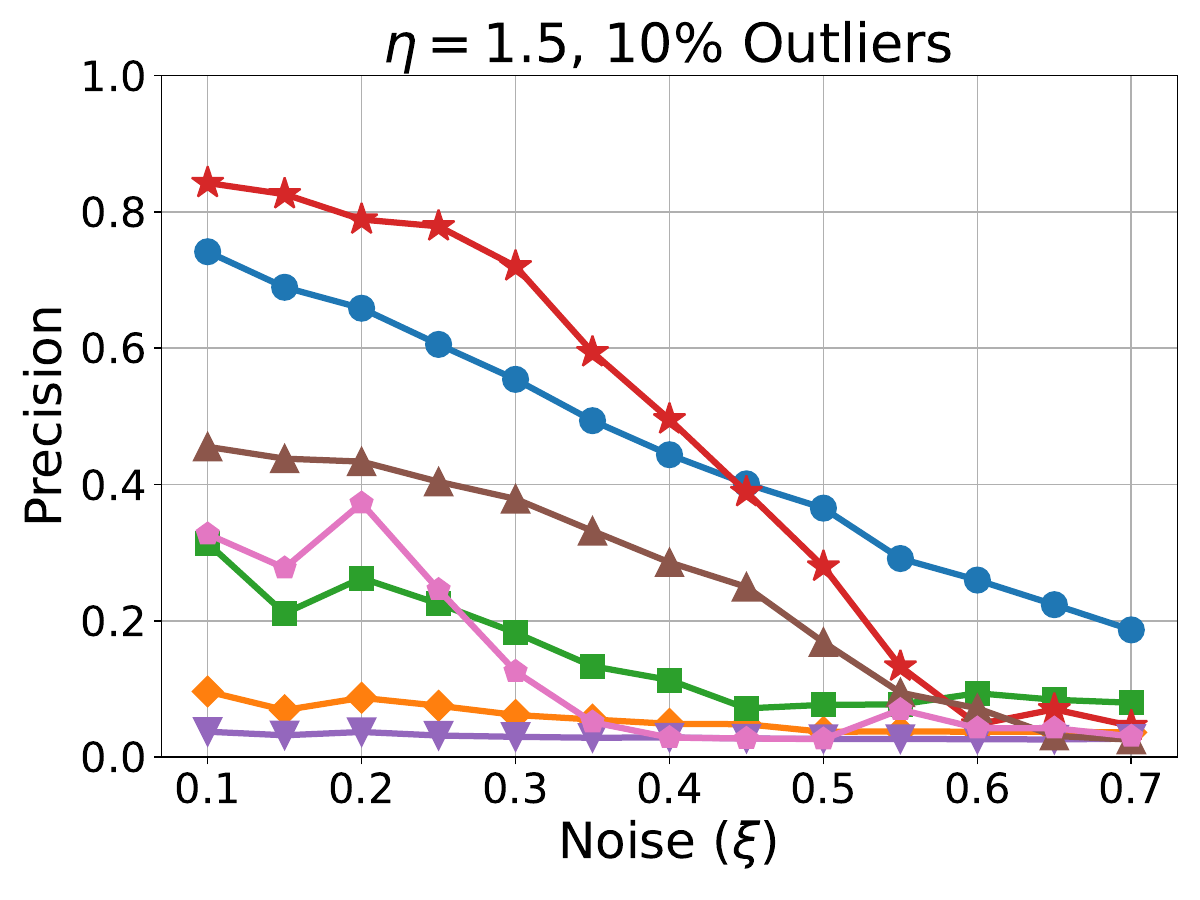}
    \includegraphics[width=0.495\linewidth]{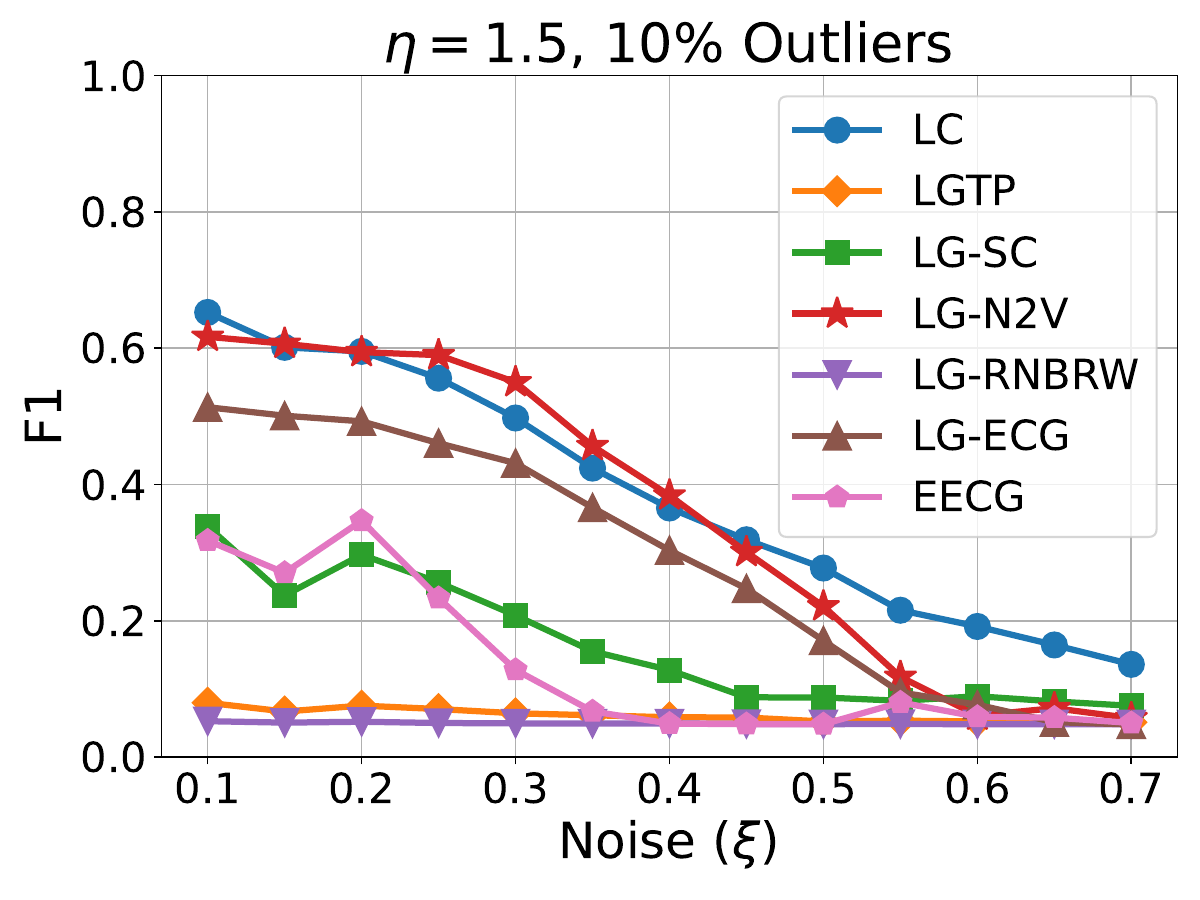}\\
    \vspace{1em}
    \includegraphics[width=0.495\linewidth]{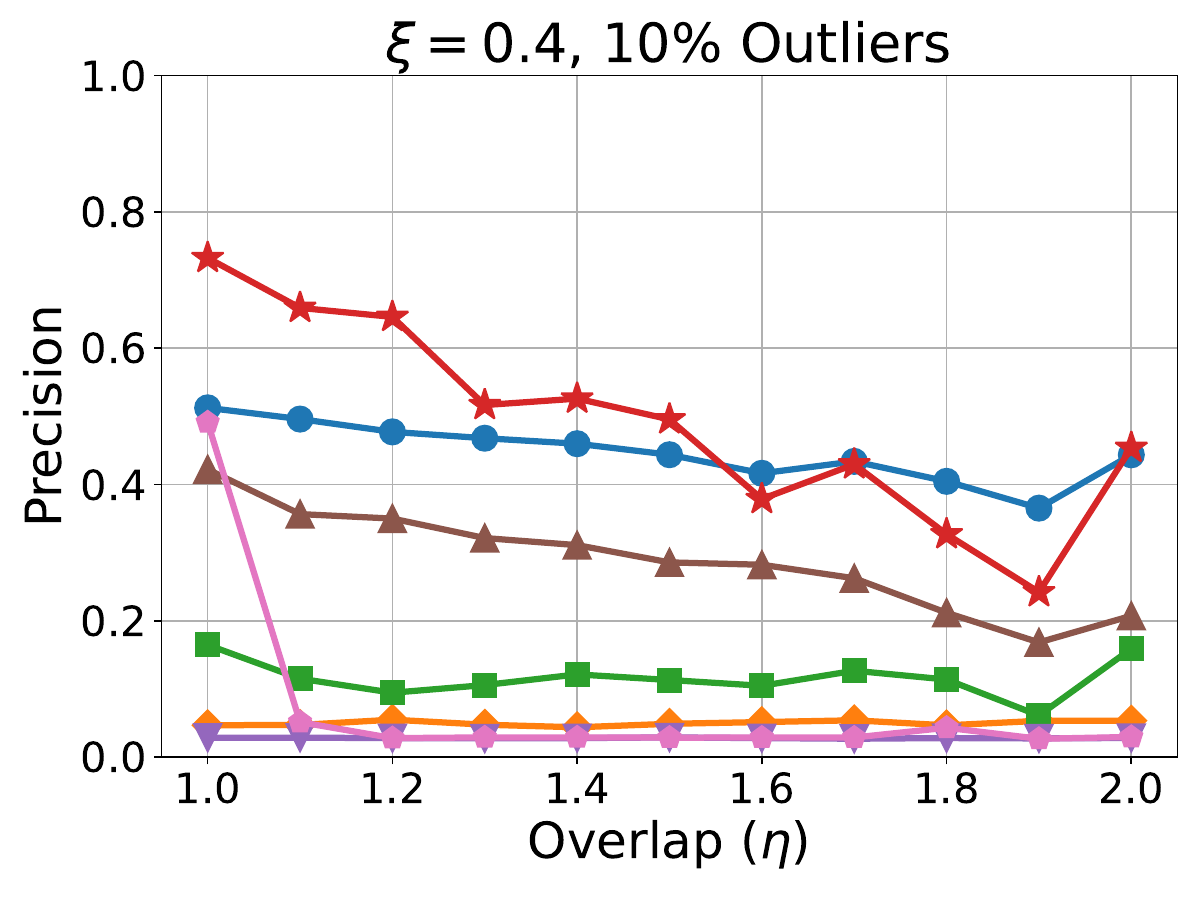}
    \includegraphics[width=0.495\linewidth]{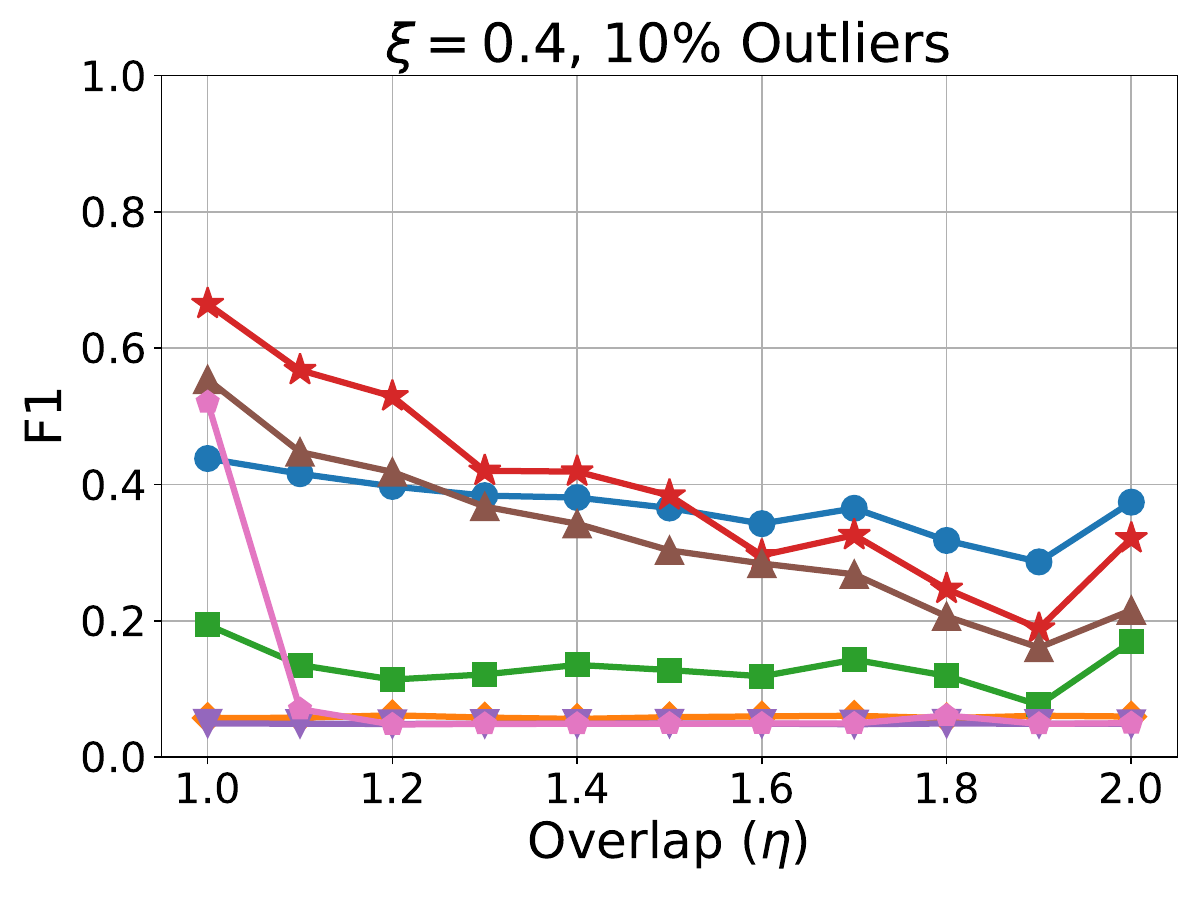}
    \caption{
        Results of the edge clustering benchmark on ABCD+o$^2$ \cite{abcdoo} graphs with $10,000$ nodes.
        We report the precision (left column) and F1 score (right column) when varying either the noise parameter $\xi$ (top row) or the amount of overlap $\eta$ (bottom row).
        The results shown are the average of each metric on $25$ random graphs produced with the given parameters.
    }
    \label{fig:abcdoo_benchmark}
\end{figure}

The performance of each edge clustering method on the synthetic ABCD+o$^2$ graphs are shown in Figure~\ref{fig:abcdoo_benchmark}, again with $10,000$ nodes and averaged over $25$ graphs.
Only link community (LC), node2vec (LG-N2V) and edge-ecg (EECG) appear to be viable option for detecting overlapping clusters, with link communities and node2vec generally out performing edge-ecg.

Finally, we apply each of the node clustering methods to the real world data-sets.
The results are shown in Table~\ref{tab:edge_clustering_results}.
Link communities (LC) is the only method with competitive performance on all four graphs.
The other method of note is Edge-ECG (EECG), which achieves almost perfect F1 on MNIST, and performs reasonably well every other graph.

\begin{table}[p]
    \centering
    \caption{
        Results from HDBSCAN cluster selection run on the edge similarity graphs.
        We set $m_s$ to $10$, $2000$, $15$, $15$ for the Football, MNIST, DBLP and Amazon graphs respectively.
        The largest precision and F1 score for each graph has been bolded.
    }
    \vspace{1em}
    \begin{tabular}{|Sc|Sc|Sc|Sc|Sc|Sc|} \hline
        Method & \# Clusters & Max Cluster Size & Coverage & Precision & F1 \\
        \hline
        \hline
        \multicolumn{6}{|Sc|}{Football} \\
        \hline
        LC & $14$ & $16$ & $1.00$ & $\mathbf{0.75}$ & $\mathbf{0.78}$ \\
        LGTP & $8$ & $105$ & $1.00$ & $0.32$ & $0.42$ \\
        LG-SC & $4$ & $98$ & $1.00$ & $0.25$ & $0.35$ \\
        LG-N2V & $11$ & $32$ & $0.98$ & $0.67$ & $0.74$ \\
        LG-ECG & $11$ & $35$ & $1.00$ & $0.46$ & $0.61$ \\
        LG-RNBRW & $3$ & $111$ & $1.00$ & $0.17$ & $0.27$ \\
        EECG & $10$ & $37$ & $1.00$ & $0.61$ & $0.72$ \\
        \hline
        \hline
        \multicolumn{6}{|Sc|}{MNIST} \\
        \hline
        LC & $19$ & $7450$ & $0.31$ & $0.94$ & $0.54$ \\
        LGTP & $5$ & $57763$ & $0.85$ & $0.16$ & $0.20$ \\
        LG-SC & $2$ & $69052$ & $0.99$ & $0.12$ & $0.19$ \\
        LG-N2V & $3$ & $45853$ & $0.94$ & $0.32$ & $0.43$ \\
        LG-ECG & $8$ & $22204$ & $1.00$ & $0.76$ & $0.82$ \\
        LG-RNBRW & $3$ & $38941$ & $0.58$ & $0.17$ & $0.23$ \\
        EECG & $10$ & $7999$ & $1.00$ & $\mathbf{0.97}$ & $\mathbf{0.97}$ \\
        \hline
        \hline
        \multicolumn{6}{|Sc|}{DBLP} \\
        \hline
        LC & $19991$ & $197$ & $0.80$ & $\mathbf{0.56}$ & $0.27$ \\
        LGTP & $20123$ & $178$ & $0.72$ & $\mathbf{0.56}$ & $\mathbf{0.29}$ \\
        LG-SC & $18$ & $314355$ & $0.99$ & $0.02$ & $0.05$ \\
        LG-N2V & $3$ & $317065$ & $1.00$ & $0.02$ & $0.05$ \\
        LG-ECG & $3$ & $317053$ & $1.00$ & $0.02$ & $0.05$ \\
        LG-RNBRW & $209$ & $306017$ & $0.97$ & $0.04$ & $0.05$ \\
        EECG & $4975$ & $21409$ & $0.94$ & $0.28$ & $0.21$ \\
        \hline
        \hline
        \multicolumn{6}{|Sc|}{Amazon} \\
        \hline
        LC & $16578$ & $550$ & $0.87$ & $\mathbf{0.87}$ & $0.46$ \\
        LGTP & $16120$ & $21599$ & $0.76$ & $0.78$ & $0.38$ \\
        LG-SC & $9182$ & $4882$ & $0.86$ & $0.80$ & $0.45$ \\
        LG-N2V & $12661$ & $697$ & $0.76$ & $0.87$ & $\mathbf{0.52}$ \\
        LG-ECG & $3069$ & $9578$ & $0.97$ & $0.76$ & $0.49$ \\
        LG-RNBRW & $5$ & $334534$ & $1.00$ & $0.16$ & $0.28$ \\
        EECG & $2482$ & $9021$ & $0.96$ & $0.76$ & $0.49$ \\
        \hline
    \end{tabular}
    \label{tab:edge_clustering_results}
\end{table}

\section{Conclusion}

We proposed applying the hierarchical single-linkage clustering method from HDBSCAN \cite{hdbscan} to several existing node and edge similarity measures to create graph clustering methods that allow for outliers and overlapping clusters.
The results on synthetic and real data suggest that several methods can perform well, and that the best performing method depends on the nature of the graph, the amount of noise, and the amount of overlap.
We believe the respectable performance indicates that hierarchical single-linkage clustering is a viable avenue for community detection with overlaps and outliers, and hope to develop improved node/edge similarity measures in future work.

\bibliographystyle{unsrt}  
\bibliography{references}  
\end{document}